\documentstyle[12pt,amsfonts]{article}

\def\half{\textstyle{\frac{1}{2}}}
\def\quarter{\textstyle{\frac{1}{4}}}

\def\H{{\cal H}}
\def\EE{{\cal E}}
\def\p{\phi}

\def\l{\lambda}

\def\S{\Sigma'}
\def\t{\textstyle}

\def\F{{\cal F}}

\def\ra{\rightarrow}
\def\tint{{\textstyle\int}}

\def\s{\hskip.08em}

\def\d{\partial}

\def\b{\begin{eqnarray}}  
\def\e{\end{eqnarray}}    
\def\bn{\begin{eqnarray}}  

\def\<{\langle}
\def\>{\rangle}

\def\no{\nonumber}

\def\k{\kappa}
\def\{{\lbrace}
\def\}{\rbrace}
\begin{document}
\title{Divergence-free Nonrenormalizable  Models}
\author{John R.~Klauder\\
Department of Physics and \\Department of Mathematics\\
University of Florida\\
Gainesville, FL 32611}

\date{}    
\maketitle
\begin{abstract}
A natural procedure is introduced to replace the traditional,
perturbatively generated counter terms to yield a formulation of
covariant, self-interacting, nonrenormalizable scalar quantum field
theories that has the added virtue of exhibiting a divergence-free
perturbation analysis. To achieve this desirable goal it is
necessary to reexamine the meaning of the free theory about which
such a perturbation takes place.
\end{abstract}
\section{Introduction}
Nonrenormalizable quantum field theories, such as $\varphi^4_n$
models, with a spacetime dimension $n\ge5$, need nontrivial counter
terms for otherwise they lead to (generalized) free theories as
shown by Aizenman \cite{aiz} and Fr\"ohlich \cite{fro}. A free
quantum theory has a trivial classical limit and so it can not
correspond to the correct quantization of the original nontrivial
classical theory. Regularized perturbation theory suggests an
unending series of distinct and ever more singular counter terms
which cannot be considered an acceptable solution. This situation
suggests that we look elsewhere for suitable counter terms, and this
paper addresses one such search.

In Sec.~2 we outline a Euclidean space lattice model for
$\varphi^4_n$ models for $n\ge5$, which includes an unconventional
counter term along with the traditional terms expected in such a
lattice formulation. In Sec.~3 we present the background for
choosing this form for the model and explain the rationale for
choosing the atypical counter term. Section 4 shows that general
correlation functions can be suitably bounded by correlation
functions at a sharp time as determined by the ground-state
distribution. In Sec.~5 we take up the question of the continuum
limit and study mass renormalization, coupling constant
renormalization, and  field strength renormalization. Here we show,
thanks to the properties of our chosen counter term, that a suitable
perturbation theory of the quartic interaction is {\it divergence
free}.  Importantly, this perturbation theory is not about the usual
free theory but about a pseudofree theory, which is a model that
contains the atypical counter term but does not include the quartic
interaction. The reason for this divergence-free character is
related to a simple idea already illustrated by idealized lattice
space integrals presented in Sec.~3. Finally, Sec.~6 offers
additional discussion and conclusions, and suggests a possible
application of the present kind of approach to other models, most
specifcally the $\varphi^4_4$ model which is perturbatively
renormalizable, but generally regarded as becoming a (generalized)
free theory when studied as the continuum limit of a conventional
lattice formulation. While our proposal seems to be analytically
challenging, there is the strong possibility that numerical Monte
Carlo methods may prove useful.

\section{Overview of the Model}
\subsection*{Preliminaries}
 The present section is devoted to a
presentation of the model, while the following sections discuss the
motivation and analysis that has led to the present formulation. We
focus on $\varphi^4_n$, $n\ge5$, models formulated as Euclidean
functional integrals; other models may possibly be treated by
analogous procedures.

We suppose Euclidean spacetime is replaced by a periodic, hypercubic
lattice with $L$ sites on an edge, $L<\infty$, and a uniform lattice
spacing of $a$, $a>0$. Let the sites be labeled by
$k=(k_0,k_1,k_2,\ldots,k_s)$, where $k_j\in {\mathbb Z}$, $k_0$
denotes the future time direction under a Wick rotation, and
$s=n-1$. We denote lattice sums (and products) over all lattice
points by $\Sigma_k$ (and $\Pi_k$), and, importantly, lattice sums
(and products) over just a spatial slice at a fixed value of $k_0$
by $\Sigma'_k$ (and $\Pi'_k$). The total number of sites is $N=L^n$,
while the number of lattice sites in a spatial slice is $N'=L^s$.

In eventually taking the continuum limit we shall do so in two
steps. First, we let $L\ra\infty$ and $a\ra0$ together so that the
full spacetime volume $V=(L\s a)^n$ remains large but finite; so too
for the spatial volume $V'=(L\s a)^s$. Second, we take the limit
that both $V$ and $V'$ diverge. In this fashion we can discuss
finite spatial volumes which would have been less convenient if we
had let $L\ra\infty$ before taking the limit $a\ra0$.

\subsection*{Lattice action}
Following aspects of the discussion in \cite{k3}, we first introduce
an important set of dimensionless constants by \b J_{k,\s
l}\equiv\frac{1}{2s+1}\s\delta_{\s k,\s l\in\{k\s\cup \s
k_{nn}\}}\;, \e where $\delta_{k,l}$ is a Kronecker delta. This
notation means that an equal weight of $1/(2s+1)$ is given to the
$2s+1$ points in the set composed of $k$ and its $2s$ nearest
neighbors in the spatial sense only; $J_{k,\s l}=0$ for all other
points in that spatial slice. {\bf [}Specifically, we define
$J_{k,\s l}=1/(2s+1)$ for the points $l=k=(k_0,k_1,k_2,\ldots,k_s)$,
$l=(k_0,k_1\pm1,k_2,\ldots,k_s)$,
$l=(k_0,k_1,k_2\pm1,\s\ldots,k_s)$,\ldots,
$l=(k_0,k_1,k_2,\ldots,k_s\pm1)$.{\bf ]} This definition implies
that $\Sigma'_l\s J_{k,\s l}=1$.

We next write the lattice action for the full theory, including the
quartic nonlinear interaction as well as the proposed counter term,
as
 \b &&\hskip-1.6em I(\p,\hbar,N)=\half{\t\sum_k} {\t\sum_{k^*}}\,(\phi_{k^*}-\phi_k)^2\,
 a^{n-2}+\half m_0^2{\t\sum_k} \phi_k^2\, a^n\no\\
 &&\hskip3cm + \l_0{\t\sum_k}\phi^4_k\,a^n+\half\s\hbar^2{\t\sum}_k\s\F_k(\p)\,a^n\,,\label{j2}\e
 where $k^*$ denotes the $n$ nearest neighbors to $k$ in
the positive sense, i.e.,
$k^*\in\{\,(k_0+1,k_1,\ldots,k_s)\s,\ldots,\s(k_0,k_1,\dots,k_s+1)\,\}$.
The last term, which represents the heart of the present procedure,
is the suggested counter term and is given (with all the following
sums over the spatial slice at fixed $k_0$) by  \b &&\F_k(\p)
\equiv\frac{1}{4}\s\bigg(\frac{N'-1}{N'}\bigg)^2\s a^{-2s}\s{\t\sum'_{\s
r,\s t}}\s\frac{J_{r,\s k}\s
  J_{t,\s k}\s \p_k^2}{[\S_l\s J_{r,\s l}\s\p^2_l]\s[\S_m\s
  J_{t,\s m}\s\p_m^2]} \no\\
  &&\hskip2.5cm-\frac{1}{2}\s\bigg(\frac{N'-1}{N'}\bigg)
  \s a^{-2s}\s{\t\sum'_{\s t}}\s\frac{J_{t,\s k}}{[\S_m\s
  J_{t,\s m}\s\p^2_m]} \no\\
  &&\hskip2.5cm+\bigg(\frac{N'-1}{N'}\bigg)
  \s a^{-2s}\s{\t\sum'_{\s t}}\s\frac{J_{t,\s k}^2\s\p_k^2}{[\S_m\s
  J_{t,\s m}\s\p^2_m]^2}\;. \e
Observe that we have included the proper dependence on $\hbar$ for
the counter term implying that its contribution  disappears in the
classical limit in which $\hbar\ra0$. It may be noticed that each of
the separate parts of the counter term scales as the {\it inverse
square} of the overall field magnitude.
 The reason
we have chosen the given counter term will be discussed in the
following sections; in Sec.~6 we even show that the counter term
may be considered to arise from a factor ordering ambiguity of the
conventional theory.

One feature of the counter term is the fact that each term involves
up to two nearest-neighbor, spatially separated lattice points. This
feature is part of the regularization in the lattice formulation of
the model. However, if a second-order phase transition is achieved
in the continuum limit, then such a regularization should still lead
to a relativistic theory in the limit.

It is important to note that as $\l_0\ra0$ and the quartic
interaction is turned off, the lattice action does {\it not} pass to
that of the usual free theory but to that of the free theory plus
the original counter term. Such a theory has been called a {\it
pseudofree theory} \cite{klau2}, and we shall show that the
interacting theory with $\l_0>0$ exhibits a divergence-free
perturbation series about the pseudofree theory. A natural argument
in favor of the pseudofree theory is given in Sec.~6.

In conventional quantum field theory, counter terms are chosen to
deal with the {\it emergence} of divergences; in the approach
adopted in this paper, the counter term is chosen to deal with the
{\it cause} of divergences. Relative to conventional treatments,
therefore, it is safe to say that using the new counter term {\it
changes everything} relative to what one normally expects based on
the usual free theory. In particular, do not look for `normal
ordering'; instead, look for `multiplicative renormalization'.

\subsection*{Generating function} One important ingredient has been
left out of the lattice action, and that is the factor $Z$
representing the field strength renormalization. We introduce this
factor most simply by adopting the following expression for the
lattice space generating function:
  \b S(h)\equiv M_0\s\int e^{\t\s Z^{-1/2}\s\Sigma_k h_k\s\p_k\s a^n/\hbar
  -I(\p,a,N)/\hbar\s}\;\Pi_k\s d\p_k\;,\label{e3} \e
  where $\{h_k\}$ determines an appropriate test sequence, and
  the normalization factor $M_0$ ensures that $S(0)=1$. By
  a field rescaling, i.e., $\p_k\ra Z^{1/2}\s\p_k $, the factor $Z$ can be
  removed from the source term and introduced into the lattice
  action; we shall have occasion to use both forms of this integral. The
  form of the generating function in terms of physical fields is given in Eq.~(\ref{h7}).

  The continuum limit will be taken as
    \b E\{h\}\equiv\lim_{a,\s L}\,S(h)\;, \e
  as $a\ra0$ and $L\ra\infty$ together such that, as discussed above, $C\equiv L\s a$ remains
  constant and finite. The argument of $E$ involves
  suitable limiting
    test functions $h_k\ra h(x)$ where $k\s a\ra x\in{\mathbb R}^n$.
    For sufficiently large $C$, it may be unnecessary to take the
    final limit $C\ra\infty$.

\section{Rationale for Counter Term}
From the lattice action it is a simple step to write down the
lattice Hamiltonian operator
  \b  &&\H \equiv -\half\s{\hbar^2}\, a^{-s}\s{\t\sum_k}'\frac{\d^2}{\d \phi_k^2}+ {\cal
  V}(\p)\no\\
  &&\hskip.47cm\equiv-\half\s{\hbar^2}\, a^{-s}\s{\t\sum_k}'\frac{\d^2}{\d \phi_k^2}+ {\cal
  V}_0(\p)+\half\s\hbar^2{\t\sum_k}'\F_k(\p)\,a^s\no  \\
  &&\hskip.47cm= -\half\s{\hbar^2}\, a^{-s}\s{\t\sum_k}'\frac{\d^2}{\d \phi_k^2}
  +\half{\t\sum'_k}{\t\sum'_{k^*}}\,(\p_{k^*}-\p_k)^2a^{s-2}+\half m_0^2{\t\sum_k}' \phi_k^2\, a^s\no\\
  &&\hskip1.5cm+\s\l_0{\t\sum_k}'\phi^4_k\,a^s+\half\s\hbar^2{\t\sum_k}'\F_k(\p)\,a^s
        -E_0\;.  \label{j7}\e
  In addition, we introduce the ground state $\Psi(\p)$ for this Hamiltonian
  operator. The constant $E_0$ is chosen so that $\Psi(\p)$ satisfies the Schr\"odinger equation
    \b \H\,\Psi(\p)=0\;, \e
    which implies that the Hamiltonian operator can also be written as
\b \H = -\frac{\hbar^2}{2}\, a^{-s}\s{\t\sum_k}'\frac{\d^2}{\d
\phi_k^2}+\frac{\hbar^2}{2}\, a^{-s}
   \s{\t\sum_k}'\frac{1}{\Psi(\p)}\s\frac{\d^2\s\Psi(\p)}{\d\phi_k^2}\;. \e
   Since the ground state $\Psi(\p)$ does not vanish, it can be
   written in the generic form
     \b \Psi(\p)=\frac{e^{\t-U(\p,a,\hbar)/2}}{D(\p)}\;,\label{j12} \e
     and thus (using the abbreviation $X,_k\equiv\d X/\d\p_k$ and the spatial summation convention)
     \b &&{\cal V}(\p)=\half\s{\hbar^2}\, a^{-s}
   \s D\,e^{U/2}\s[\s D^{-1}\s e^{-U/2}\s],_{kk}\label{j13}\\
       &&\hskip1cm=\half\s{\hbar^2}\, a^{-s}\s[\,\quarter\s U^2,_k-\half\s U,_{kk}+D^{-1}\s
       U,_k\s D,_k+2\s D^{-2}\s D^2_k-D^{-1}\s D,_{kk}\s]\;. \no\e
     We insist that the atypical counter term $\F_k(\p)$ is determined
     by the denominator $D$ alone by requiring that
       \b \hskip-.2cm\half\s\hbar^2{\t\sum_k}'\F_k(\p)\,a^s\equiv\half{\hbar^2}\,a^{-s}\,D\,
       D^{-1},_{kk}=\half{\hbar^2}\,a^{-s}[2\s D^{-2}\s D^2,_k-D^{-1}\s D,_{kk}\s]\;. \label{j14}\e
       There are multiple solutions to this
       equation which lead to ground state functions that are locally square integrable
       near the origin in field space. However, the only solution consistent with a
       nowhere vanishing ground state is given (up to an overall factor) by
        \b D(\p)=\Pi'_k\,[\s\Sigma'_l\,J_{k,\s l}\,\p_l^2\s]^{(N'-1)/4N'}\;.
       \e
       In point of fact, $D$  was chosen {\it first}, and
       the counter term was then {\it derived}  from $D$ by this
       very differential equation.
       Why we have chosen this specific form for $D$
       is discussed below.

       The ground state $\Psi(\p)$ leads to the probability density
       \b  \Psi(\p)^2 \equiv K\,\frac{e^{\t-U(\p,a,\hbar)}}{\Pi'_k\,
       [\Sigma'_l\,J_{k,\s l}\,\p_l^2]^{(N'-1)/2N'}}\;, \e
       where $K$ accounts for normalization of this expression
       given the additional assumption that $U(0,a,\hbar)=0$. The normalization
       integral itself then reads
        \b K\,\int \frac{e^{\t-U(\p,a,\hbar)}}{\Pi'_k\,
       [\Sigma'_l\,J_{k,\s
       l}\,\p_l^2]^{(N'-1)/2N'}}\;\Pi'_k\,d\p_k=1\;.\label{z7}
       \e

       Before commenting on this integral further, we wish to
       discuss several simpler integrals.

  \subsection*{A discussion of many-dimensional integrals}
        Consider the family of Gaussian integrals given (at some fixed $k_0$) by
\b I_G(2p)\equiv \int [\s\Sigma'_k\s\p_k^2\s]^p e^{\t
-A\s\Sigma'_k\s\p_k^2}\;\Pi'_k\,d\p_k\;, \e
         where $p\in\{0,1,2,3,\ldots\s\}$, and we assume that $A$ is of `normal size', e.g.,
         $0.1<A<10$.
         Although these integrals can be evaluated explicitly, we prefer to
         study the qualitative behavior of such integrals for {\it large values
         of} $N'$, i.e., when $N'\gg1$. For this purpose it is highly instructive
         to introduce {\it hyper-spherical coordinates} \cite{klau,iande}
         defined by
        \b &&\p_k\equiv\k\s\eta_k\;,\hskip.5cm
        0\le\k<\infty\;,\hskip.5cm -1\le\eta_k\le1\;, \no\\
        &&\hskip1.4cm\S_k\eta_k^2\equiv1\;,\hskip.5cm \S_k\p_k^2\equiv\k^2\;. \e
Here $\kappa$ acts as a hyper-radius field variable and the
          $\{\eta_k\}$ variables constitute an $N'$-dimensional direction field.
Note well that $\k\equiv\sqrt{\Sigma'_k\s\p_k^2}$ is the `radius' of
{\it all the field
        variables in a given
        spatial slice of the lattice at some fixed $k_0$}. In
        terms of these variables, it follows that
          \b  I_G(2p)=2\int[\s\kappa^2\s]^p\,e^{\t-A\s\kappa^2}\,\kappa^{N'-1}\s
          d\kappa\,\delta(1-\Sigma'_k\eta_k^2)\,\Pi'_k\,d\eta_k\;.\e
          Observe that the integrand depends on the radius $\kappa$,
          but it
          does not depend on the angular variables $\{\eta_k\}$. For very large $N'$,
          the integral over $\kappa$ can be studied by steepest descent
          methods. To leading order, the integrand is narrowly peaked at
          values of $\kappa\simeq (N'/2A)^{1/2}$, namely at large
          values of $\kappa$. As a consequence, for each
          value of $A$, the integrand is supported on a
          {\it disjoint} set of $\k\s$ as $N'\ra\infty$. This well-known fact \cite{hida} leads to
          divergences in perturbation calculations. For example, let
          us study
            \b I^\star_G(2)=\int[\S_k\p_k^2]\s e^{\t -A^\star\S_k\p_k^2}\,\Pi'_k d\p_k\;,\e
            which is the same type of Gaussian integral
            for a different value of $A$. For this study, we introduce the perturbation series
            \b I^\star_G(2)=I_G(2)-\Delta A\,I_G(4)+\half(\Delta
            A)^2\,I_G(6)-\cdots\;, \e
            where $\Delta A\equiv A^\star-A$. Since
            $I_G(2p)/I_G(2)=O(N'^{(p-1)})$, this series exhibits
            {\it divergences} as $N'\ra\infty$. It is not difficult
            to convince oneself that such divergences are quite
            analogous to those that appear in quantum field theory;
            see \cite{klau} for further examples of this sort.

            If the factor $\kappa^{(N'-1)}$ is removed from $I_G(2p)$, we are
            led to consider
          \b  I'_G(2p)=2\int[\s\kappa^2\s]^p\,e^{\t-A\s\kappa^2}\,
          d\kappa\,\delta(1-\Sigma'_k\eta_k^2)\,\Pi'_k\,d\eta_k\;.\e
       Now, the integrand is broadly supported and no longer favors large $\kappa$
       values. Consequently, a
       perturbation series evaluation of
          \b &&I'^\star_G(2)=2\int\s\kappa^2\s
          e^{\t-A^\star\kappa^2}\,d\kappa\,\delta(1-\Sigma'_k\eta_k^2)\,\Pi'_k\,d\eta_k\no\\
               &&\hskip1.1cm =I'_G(2)-\Delta A\,I'_G(4)+\half(\Delta
            A)^2\,I'_G(6)-\cdots \e
            exhibits no divergences since
            $I'_G(2p)/I'_G(2)=O(N'^{\, 0})$.

          Unlike the original integrals over $\kappa$, integrals
          over the $ \{\eta_k\}$ direction field variables cannot diverge
          under normal circumstances since each variable satisfies
          $-1\le\eta_k\le1$. We will encounter such
          variables in the denominator of (\ref{z7}); however, the form of
          that denominator
          has been specifically chosen to ensure that such integrals converge
          near zero for all $N'<\infty$.

\subsection*{Relevance for the ground-state distribution}
           The normalization integral for
           the ground-state distribution, Eq.~(\ref{z7}), expressed
            in terms of  hyper-spherical
       coordinates, becomes
       \b 2\s K\int \frac{e^{\t-U(\kappa\s\eta,a,\hbar)}}{\Pi'_k\,
       [\s\S_l J_{k,l}\s\eta^2_l\s]^{(N'-1)/2N'}}\;d\kappa\;
       \delta(1-\Sigma'_k\eta_k^2)\,\Pi'_k\s d\eta_k=1\;.\e
       Note well that the absence of the factor $\kappa^{(N'-1)}$ in this
       expression is a direct result of the counter term in
       the lattice Hamiltonian, which in turn gave rise to the denominator factor
       $D$ in the ground state $\Psi(\p)$. Just like the elementary
       examples in which $\k^{(N'-1)}$ was  artificially removed,
       there is no peaking of the integrand in $\k$ due to that factor.
       For integrals such as
         \b K\int [\Sigma'_k\s\p_k^{2}]^p\, \frac{e^{\t-U(\p,a,\hbar)}}{\Pi'_k\,
       [\s\S_l J_{k,l}\p_l^2\s]^{(N'-1)/2N'}}\;\Pi'_k\,d\p_k \e
       it is clear that the $\k$-dependence of the integrand is most likely
       spread rather broadly; this conclusion would be false if the
       factor $\k^{(N'-1)}$ had not been canceled by part of the term $D^{\,2}$.

 \section{Correlation Functions and their Bounds}
  In this section, following \cite{k3}, we wish to show that the
  full spacetime correlation functions can be
  controlled by their sharp-time behavior along with a suitable
  choice of test sequences.

   Let the notation
    \b  \p_u\equiv \Sigma_k u_k\s\p_k\,a^n \e
denote the full spacetime summation over all lattice sites where
$\{u_k\}$ denotes a suitable test sequence. We also separate out the
temporal part of this sum in the manner
   \b \p_u\equiv\Sigma_{k_0}a \,\p_{u'}\equiv
\Sigma_{k_0}\s a\,\S_k u_k\s\p_k\,a^s\;. \e Observe that the
notation $\p_{u'}$ (with the prime) implies a summation {\it only}
over  the spatial lattice points for a fixed (and implicit) value of
the temporal lattice value $k_0$.

Let the notation $\<(\s\cdot\s)\>$ denote full spacetime averages
with respect to the field distribution determined by the lattice
action, and then let us consider full spacetime correlation
functions such as
 \b  \hskip.2cm\<\s\p_{u^{(1)}}\s\p_{u^{(2)}}\s\cdots\s\p_{u^{(2q)}}\s\>
 = \Sigma_{k_0^{(1)},k_0^{(2)},\s\dots\s,k_0^{(2q)} }\,a^{2q}\,\<\s\p_{u'^{(1)}}
\s\p_{u'^{(2)}}\s\cdots\s\p_{u'^{(2q)}}\s\> \;,
\e
where $q\ge1$ and the expectation on the right-hand side is over products of
fixed-time summed fields, $\p_{u'}$, for possibly different times,
which are then summed over their separate times.
 All odd correlation functions are assumed
    to vanish, and furthermore, $\<\s1\s\>=1$ in this normalized
    spacetime lattice field distribution.
It is also clear that
  \b  |\<\s\p_{u^{(1)}}\s\p_{u^{(2)}}\s\cdots\s\p_{u^{(2q)}}\s\>|
 \le \Sigma_{k_0^{(1)},k_0^{(2)},\s\dots\s,k_0^{(2q)} }\,a^{2q}\,|\<\s\p_{u'^{(1)}}
\s\p_{u'^{(2)}}\s\cdots\s\p_{u'^{(2q)}}\s\>| \;.  \e At this point
we turn our attention toward the spatial sums alone.

    We appeal to straightforward inequalities of the general
    form
      \b \s\<A\s B\>^{2}\le \<\s A^2\s\>\s\<\s B^2\s\>\;.  \e
   In particular, it follows that
    \b
    \<\s\p_{u'^{(1)}}\s\p_{u'^{(2)}}\s\p_{u'^{(3)}}\s\p_{u'^{(4)}}\>^2
     \le\<\s\p^2_{u'^{(1)}}\s\p^2_{u'^{(2)}}\>\s\<\s\p^2_{u'^{(3)}}\s\p^2_{u'^{(4)}}\>\;,
     \e
     and, in turn, that
      \b &&\hskip-1cm\<\s\p_{u'^{(1)}}\s\p_{u'^{(2)}}\s\p_{u'^{(3)}}\s\p_{u'^{(4)}}\>^4
      \le\<\s\p^2_{u'^{(1)}}\s\p^2_{u'^{(2)}}\>^2\s\<\s\p^2_{u'^{(3)}}\s\p^2_{u'^{(4)}}\>^2\no\\
      &&\hskip3.2cm\le\<\s\p^4_{u'^{(1)}}\>\s\<\p^4_{u'^{(2)}}\>\s\<\s\p^4_{u'^{(3)}}\>\s\<\p^4_{u'^{(4)}}\>\;.
      \e
      By a similar argument, it follows that
      \b
      |\<\s\p_{u'^{(1)}}\s\p_{u'^{(2)}}\s\cdots\s\p_{u'^{(2q)}}\s\>|
      \le \Pi_{j=1}^{2q}\,[\s\<\p^{2q}_{u'^{(j)}}\>]^{1/2q}\;, \e
      which has bounded any particular mixture of spatial correlation functions at
      possibly different times,
      by a suitable product of higher-power expectations each of
      which involves field values ranging over a spatial level, all at a single
      fixed lattice time.
     By time translation invariance of the various single time correlation functions
      we can assert that
        \b \<\s \p^{2r}_{u'^{(j)}}\s\>\;, \e
        which is defined at time $k_0^{(j)}$, is actually independent of
        the time and, therefore, the result could be calculated at any
        fixed time. In particular, we can express such correlation functions as
        \b  \<\s\p^{2q}_{u'}\s\>=\int
        \p^{2q}_{u'}\,\Psi(\p)^2\,\Pi'_k\,d\p_k \;.\e

        Thus we see that a bound on full spacetime correlation functions may be given
        in terms of sharp-time correlation functions in the ground-state distribution.

\section{The Continuum Limit}
      Before focusing on the limit $a\ra0$ and $L\ra\infty$, let us
      note some important facts about ground-state averages of the direction
      field variables $\{\eta_k\}$. First, we assume that such averages
      have two important symmetries: (1) averages of an odd number
      of $\eta_k$ variables vanish, i.e.,
      \b \<\eta_{k_1}\cdots\eta_{k_{2p+1}}\>=0\;, \e
      and (2) such averages are invariant under any spacetime translation, i.e.,
    \b
    \<\eta_{k_1}\cdots\eta_{k_{2p}}\>=\<\eta_{k_1+l}\cdots\eta_{k_{2p}+l}\>\;\e
    for any $l\in{\mathbb Z}^n$ due to a similar translational
    invariance of the lattice Hamiltonian. Second, we note that
    for any ground-state distribution, it is
    necessary that $\<\s\eta_k^2\s\>=1/N'$
for the simple reason that $\Sigma'_k\s\eta_k^2=1$. Hence,
       $|\<\eta_k\s\eta_l\>|\le1/N'$ as follows from the Schwarz
       inequality. Since $\<\s[\s\Sigma'_k\s\eta_k^2\s]^2\>=1$, it
       follows that $\<\s\eta_k^2\s\eta_l^2\s\>=O(1/N'^{2})$. Indeed,
       similar arguments show that for any ground-state distribution
         \b  \<\eta_{k_1}\cdots\eta_{k_{2p}}\>=O(1/N'^{p})\;, \e
         which will be useful in the sequel.

         Next, we choose to study the pseudofree model, namely, when
         the coupling constant $\l_0=0$. This is as close as we can
         get to the free model itself. Unfortunately, we cannot
         solve Eq.~(\ref{j13}) when $\cal V$ has the desired form.
         The best we can do is choose a form for $U(\p,a,\hbar)$ in
         (\ref{j12}) that leads to an {\it approximate form} of
        the pseudofree model. In particular, we
         choose
           \b U(\p,a,\hbar)=(1/\hbar)\s\Sigma'_{k,l}\,\p_k\s
           A_{k-l}\s\p_l\,a^{2s}\;. \label{f1}\e
         This expression is taken to be the form for
         $U$ for the free model as if there was no counter term and
         consequently $D$ was replaced by $1$. This ensures us that
         the potential ${\cal V}(\p)$ that follows from (\ref{j13})
         agrees with the desired free model to leading order in
         $\hbar$. Specifically, with the given choice for $U$ and
         $D$, it follows that
         \b &&\hskip-.5cm{\cal V}(\p)=\half\Sigma'_{k,l,m}\,\p_k\s A_{k-l}\s
         A_{l-m}\s \p_m\,a^{3s}-\half\s\hbar\s A_0\,N'\s a^s +\half\s\hbar^2\s\Sigma'_k\s\F_k(\p)\;a^s\no\\
         &&\hskip1cm+\hbar\s[(N'-1)/4N']\s\Sigma'_{k,r,m}\s J_{r,k}\s\p_k\s
         A_{k-m}\s \p_m/[\s\S_lJ_{r,l}\s\p_l^2\s]\,a^{2s}
         \;. \label{f2}\e
         We choose the matrix $A_{k-l}$ so that the
         first term in (\ref{f2}) yields the desired gradient and
         mass terms in the Hamiltonian expression (\ref{j7}).
         In particular, we note that to match the quadratic,
         spatial lattice-derivative terms in
         the Hamiltonian (typically the most singular of the quadratic terms),
         we can do so by choosing the elements of
         the matrix $A_{k-l}=O(a^{-(s+1)})$.
         Although this leads to only an approximate form for the
         pseudofree model, it is sufficient for our
          limited purpose at
         present, namely, to determine the field strength renormalization
          constant $Z$.

\subsubsection*{Field strength renormalization}
         We now take up the question of the sharp-time averages given by
         \b \int Z^{-p}\,[\Sigma'_k h_k\s\p_k\,a^s]^{2p}\,\Psi(\p)^2\,\Pi'_k\s
         d\p_k\;, \e
         where $Z$ denotes the field strength renormalization factor and
         $\{h_k\}$ represents a suitable spatial test sequence.
         These are exactly the kinds of expression that should become
         well behaved in the continuum limit for a proper choice of
         $Z$. Thus, we are led to consider
         \b &&\hskip-.8cm K\int Z^{-p}\,[\Sigma'_k
         h_k\s\p_k\,a^s]^{2p}\,\frac{e^{\t-\Sigma'_{k,l}\s\p_k\s A_{k-l}\s\p_l\,
         a^{2s}/\hbar}}{\Pi'_k[\s\S_lJ_{k,l}\p_l^2\s]^{(N'-1)/2N'}}\,\Pi'_k\s
         d\p_k\\
         &&\hskip-.7cm=2\s K\int Z^{-p}\s\k^{2p}\,[\Sigma'_k h_k\s\eta_k\,a^s]^{2p}\,
         \frac{e^{\t-\k^2\s\Sigma'_{k,l}\s\eta_k\s A_{k-l}\s\eta_l\,a^{2s}/\hbar}}
         {\Pi'_k[\s\S_l J_{k,l}\s\eta^2_l\s]^{(N'-1)/2N'}}\,d\k\,\delta(1-\Sigma'_k\eta_k^2)
         \,\Pi'_k\s d\eta_k\;. \label{f5} \no\e

         Our  goal is to use this integral to determine a value for
         the field strength renormalization constant $Z$.
         To estimate this integral we first replace two
         factors with $\eta$ variables by their appropriate
         averages. In particular, the expression in the exponent
         is estimated by
          \b \k^2\s\Sigma'_{k,l}\s\eta_k\s
          A_{k-l}\s\eta_l\,a^{2s}\simeq\k^2\s\Sigma'_{k,l}\s N'^{\,-1}
          A_{k-l}\,a^{2s}\propto \k^2\s N'\s a^{2s}\s a^{-(s+1)}\;, \e
          and the expression in the integrand is estimated by
          \b [\Sigma'_k h_k\s\eta_k\,a^s]^{2p}\simeq\s
          N'^{\,-p}\,[\Sigma'_k h_k\,a^{s}]^{2p}\;. \e
         The integral over $\k$ is then estimated by first rescaling the variable
         $\k^2\ra\k^2/(N'\s a^{s-1})$, which then leads to an overall integral-estimate proportional
          to
         \b  Z^{-p}\,[N'\s a^{s-1}]^{\,-p}\,N'^{-p}\,[\Sigma'_k h_k\,a^{s}]^{2p}\;.\e
         Finally, for this result to be meaningful in the continuum limit,
         we are led to choose $Z=N'^{\,-2}\s a^{-(s-1)}$. However, $Z$ must
         be dimensionless, so we introduce a fixed positive quantity
         $q$ with dimensions of an inverse length, which allows us to
         set
          \b Z=N'^{\,-2}\s (q\s a)^{-(s-1)}\;. \e
          This is a fundamental and important relation in our analysis.
 \subsubsection*{Mass renormalization}
With notation where $\<(\cdot)\>$ denotes a full spacetime
lattice-space average based on the lattice action for the pseudofree
theory, an expansion of the mass term leads to a series of terms of
the form
  \b \<\s [\s  m_0^2\,\Sigma_k \p_k^2 a^n\s]^p\s\>\;, \e
  which in turn can be expressed as
    \b
    m^{2p}_0\,\s\Sigma_{{k_0^{(1)}},\s{k_0^{(2)}},\ldots,\s{k_0^{(p)}}}\,a^p\,
\<\s[\s\S_{k^{(1)}}\p^2_{k^{(1)}}\s
a^s]\s[\s\S_{k^{(2)}}\p^2_{k^{(2)}}\s
a^s]\cdots\s[\S_{k^{(p)}}\s\p^2_{k^{(p)}}\s a^s]\s\>\;. \e Based on
the inequality
  \b \<\s \Pi_{j=1}^p\s A_j\s\>\le \Pi_{j=1}^p\s\<\s
  A_j^p\s\>^{1/p}\;, \e
  valid when $A_j\ge0$ for all $j$, it follows that
  \b &&\hskip-1cm\<\s [\s  m_0^2\,\Sigma_k \p_k^2 a^n\s]^p\s\>\le
  m^{2p}_0\,\s\Sigma_{{k_0^{(1)}},\s{k_0^{(2)}},\ldots,\s{k_0^{(p)}}}\,a^p\no\\
&&\times\{\<\s[\s\S_{k^{(1)}}\p^2_{k^{(1)}}\s
a^s]^p\s\>\s\<\s[\s\S_{k^{(2)}}\p^2_{k^{(2)}}\s
a^s]^p\s\>\cdots\<\s[\S_{k^{(p)}}\s\p^2_{k^{(p)}}\s
a^s]^p\s\>\}^{1/p}\;.\e This leads us to consider \b
  &&\<\s[\s m_0^2\s\S_{k}\p^2_{k}\s\s
a^s]^p\s\>=2\s\s m_0^{2p}\s a^{sp}\s K\int \k^{2p}\,
\frac{e^{\t-\k^2\S_{k,l}\eta_k\s
A_{k-l}\s\eta_l\,a^{2s}}}{\Pi_k[\s\S_l J_{k,l}\s\eta^2_l\s]^{(N'-1)/2N'}}\,d\k\no\\
&&\hskip4cm\times\delta(1-\S_k\eta_k^2)\,\Pi\s d\eta_k\,, \e which,
in the manner used previously, can be estimated as
  \b \<\s[\s m_0^2\s\S_{k}\p^2_{k}\s\s a^s]^p\s\>\propto \,\frac{m_0^{2p}\,a^{sp}}{[N'\s a^{(s-1)}]^p}\;. \e
To make sense in the continuum limit, this leads us to identify
  \b m_0^2=N'\s(q\s a)^{-1}\,m^2\;,\e
  with $m^2$ being the physical mass. Moreover, it is noteworthy that
    \b Z\s m_0^2=[\s N'^{\,-2}\s (q\s a)^{-(s-1)}]\,[\s N'\s(q\s
    a)^{-1}\s]\,m^2=[\s N'\s (q\s a)^s\s]^{-1}\s m^2\;,\label{x2} \e
    which for a finite spatial volume $V'=N'\s a^s$ leads to a
    finite nonzero result for $Z\s m_0^2$.

\subsubsection*{Coupling constant renormalization}
We repeat the previous calculation for an expansion of the quartic
interaction term about the pseudofree theory. This leads us to
consider terms of the form
 \b \<\s [\s  \l_0\,\Sigma_k \p_k^4 a^n\s]^p\s\>\;, \e
  which in turn can be expressed as
    \b
    \l^{p}_0\,\s\Sigma_{{k_0^{(1)}},\s{k_0^{(2)}},\ldots,\s{k_0^{(p)}}}\,a^p\,
\<\s[\s\S_{k^{(1)}}\p^4_{k^{(1)}}\s
a^s]\s[\s\S_{k^{(2)}}\p^4_{k^{(2)}}\s
a^s]\cdots\s[\S_{k^{(p)}}\s\p^4_{k^{(p)}}\s a^s]\s\> \e and bounded
by \b &&\hskip-1cm\<\s [\s  \l_0\,\Sigma_k \p_k^4 a^n\s]^p\s\>\le
  \l^{p}_0\,\s\Sigma_{{k_0^{(1)}},\s{k_0^{(2)}},\ldots,\s{k_0^{(p)}}}\,a^p\no\\
&&\times\{\<\s[\s\S_{k^{(1)}}\p^4_{k^{(1)}}\s
a^s]^p\s\>\s\<\s[\s\S_{k^{(2)}}\p^4_{k^{(2)}}\s
a^s]^p\s\>\cdots\<\s[\S_{k^{(p)}}\s\p^4_{k^{(p)}}\s
a^s]^p\s\>\}^{1/p}\;.\e This leads us to consider \b
  &&\<\s[\s \l_0\s\S_{k}\p^4_{k}\s\s
a^s]^p\s\>=2\s\s \l_0^{p}\, a^{sp}\s K\int
\k^{4p}\,[\s\S_{k}\s\eta^4_{k}\s]^p\,
\frac{e^{\t-\k^2\S_{k,l}\eta_k\s
A_{k-l}\s\eta_l\,a^{2s}}}{\Pi_k|[\s\S_l J_{k,l}\s\eta^2_l\s]^{(N'-1)/2N'}}\no\\
&&\hskip4cm\times \,d\k\,\delta(1-\S_k\eta_k^2)\,\Pi\s d\eta_k\,, \e
which, in the manner used previously, can be estimated as
  \b \<\s[\s \l_0\s\S_{k}\p^4_{k}\s a^s]^p\s\>\propto \frac{\l_0^{p}\,N'^{\,-p}\s a^{sp}}
  {[N'\s a^{(s-1)}]^{2p}}\;, \e
and to make sense in the continuum limit leads us to identify
  \b \l_0=N'^{\,3}\s(q\s a)^{s-2}\,\l\;,\e
  with $\l$ being the physical coupling constant. Moreover, it is noteworthy that
    \b Z^2\s \l_0=[\s N'^{\,-4}\s (q\s a)^{-2(s-1)}]\,[\s N'^{\,3}\s(q\s
    a)^{s-2}\s]\,\l=[\s N'\s (q\s a)^s\s]^{-1}\s \l\;, \label{x4}\e
    which for a finite spatial volume $V'=N'\s a^s$ leads to a
    finite nonzero result for $Z^2\s \l_0$.

\subsection*{Physical version of the generating function}
Based on the previous analysis we are led to reformulate the
expression for the lattice space generating function (\ref{e3}). We
first make a change of integration variables such that $\p_k\ra
Z^{1/2}\s\p_k$ leading to the expression
 \b S(h)\equiv M\s\int e^{\t \Sigma_k h_k\s\p_k\s a^n/\hbar
  -I(Z^{1/2}\s\p,a,N)/\hbar\s}\;\Pi_k\s d\p_k\;,\label{h3} \e
  where any constant Jacobian factor has been absorbed into a change of the overall
  normalization factor from $M_0$ to $M$. Finally, we introduce the
  explicit form for the lattice action from (\ref{j2}) into
  (\ref{h3}) to yield
  \b &&S(h)=M\int\exp\{\s\Sigma h_k\s\p_k\,a^n/\hbar\no\\ &&\hskip1.5cm-\s
 \half\,[N'^{\,2}\s(q\s a)^{(s-1)}]^{-1}
 {\t\sum_k} {\t\sum_{k^*}}\,(\phi_{k^*}-\phi_k)^2\,
 a^{n-2}/\hbar\no\\&&\hskip1.5cm
 -\half [\s N'\s (q\s a)^s\s]^{-1}\s m^2\,{\t\sum_k} \phi_k^2\,
 a^n/\hbar\s-\s [\s N'\s (q\s a)^s\s
 ]^{-1}\s\l\,{\t\sum_k}\phi^4_k\,a^n/\hbar\no\\
 &&\hskip1.5cm-\s\half\s\hbar^2\s[\s N'^{\,2}\s(q\s a)^{(s-1)}\s]\s
 {\t\sum_k}\F_k(\p)\,a^n/\hbar\s\}\;\Pi_k\s d\p_k\;. \label{h7} \e
 This expression contains a formulation of the lattice space
 generating function expressed in terms of physical fields and constants.

 \subsubsection*{Commentary}
In our final expression above there are several noteworthy points to
be made.
 In a finite spatial volume $V'=N'\s a^s$ -- which due to our hypercubic
assumption for spacetime implies a finite spacetime volume $V=N\s
a^n$ -- the coefficients of the physical mass $m$ and the physical
coupling constant $\l$ are both finite and nonzero.
 We have shown earlier in this section that perturbation
in both the quadratic mass term and the quartic nonlinear action
leads to a series which is term-by-term finite when perturbed about
the pseudofree theory. If the finite spatial volume is taken large
enough (e.g., Milky Way sized), then a divergence-free perturbation
series is established. In other words, the introduction of the
unusual counter term has resolved any issues with typical
ultraviolet divergences (and it is noteworthy that it has not been
necessary to maintain an ultraviolet cutoff to achieve this goal).

{\bf [Remark:} Any theory exhibits infinite volume divergences for
questions of a stationary nature. For example, such divergences even
arise already when $n=1$ and we deal with time alone, as for example
with a conventional, stationary Ornstein-Uhlenbeck (O-U) process
$U(t)\equiv 2^{-1/2}\,e^{\t-t}\,W(e^{\t 2t})$, $-\infty<t<\infty$,
where $W(\tau)$, $0\le\tau<\infty$, denotes a standard, Gaussian,
Wiener process for which $W(0)=0$, ${\bf E}(W(\tau))=0$, and ${\bf
E}(\s W(\tau_1)\s W(\tau_2))=\min(\tau_1,\tau_2)$, where ${\bf E}$
denotes ensemble average. Although the O-U paths are concentrated on
bounded, continuous paths, it nevertheless follows that \b {\bf
E}(\tint_{-\infty}^\infty U(t)^2\,dt)=\infty \e due to the
stationarity of the process.{\bf ]}

Unlike the quadratic mass and quartic interaction terms, the
coefficients of the derivative terms and the inverse-square field
counter term are inverse to one another and do not have finite
nonzero limits when $L\ra\infty$ and $a\ra0$ such that the spatial
volume $V'=(L\s a)^s$ is finite. This aspect is not unexpected since
(i) the coefficient of the counter term is intimately linked to that
of the derivative term so that the field power that appears in the
denominator factor $D^2$ in the ground state distribution is
$(N'-1)/2N'$, and (ii) this fact leads to a significant
redistribution of
 probability toward the origin of field space, which then requires
 an asymptotically small $Z$ factor to reestablish reasonable field
 averages.

\subsection*{Numerical studies}
 In a certain sense, the most basic pseudofree model lacks both the
 quartic coupling and the mass term. This leads to the idealized (I)
 model described by
   \b &&S_I(h)\equiv M\int\exp\{\s\Sigma h_k\s\p_k\,a^n/\hbar\no\\ &&\hskip1.5cm-\s
 \half\,[N'^{\,2}\s(q\s a)^{(s-1)}]^{-1}
 {\t\sum_k} {\t\sum_{k^*}}\,(\phi_{k^*}-\phi_k)^2\,
 a^{n-2}/\hbar\no\\
 &&\hskip1.5cm-\s\half\s\hbar^2\s[\s N'^{\,2}\s(q\s a)^{(s-1)}\s]\s
 {\t\sum_k}\F_k(\p)\,a^n/\hbar\s\}\;\Pi_k\s d\p_k\;. \label{h271} \e
 If we combine the factors of $a$, as well as transform the fields
 $\p_k$ to remove both $\hbar$ and $q$ from the idealized action,
 it follows that
 \b &&\hskip-.5cm S_I(h)=M\int \exp\{\s\Sigma {\tilde h}_k\s\p_k\,a^n \no\\ &&\hskip1.1cm-\s
 \half\s N'^{\,-2}\s
 {\t\sum_k} {\t\sum_{k^*}}\,(\phi_{k^*}-\phi_k)^2 -\s\half\s N'^{\,2}\s
 {\t\sum_k}\EE_k(\p)\s\}\;\Pi_k\s d\p_k\;, \label{h77} \e
where ${\tilde h}_k\equiv q^{(s-1)/2}\s h_k/\hbar^{1/2}$ and
    \b \EE_k(\p)\equiv a^{2s}\s\F_k(\p)\;;\e
note that $\EE_k(\p)$ is dimensionless apart from any that may arise
from its inverse-square field dependence.
    One
    last transformation, in which $\p_k\ra N'\s \p_k$, leads to the
    expression
    \b &&\hskip-.5cm S_I(h)=M\int \exp\{\s\Sigma {\tilde h}_k\s\p_k\,N'\,a^n\no\\ &&\hskip1.1cm-\s
 \half\s
 {\t\sum_k} {\t\sum_{k^*}}\,(\phi_{k^*}-\phi_k)^2\s-\s\half \s
 {\t\sum_k}\EE_k(\p)\s\}\;\Pi_k\s d\p_k\;; \label{h97} \e
in the above expressions we have used the same symbol ($M$) for the
normalization factor even though it has absorbed different Jacobian
factors. This last version may be useful for numerical studies of
this basic pseudofree model. As argued in Sec.~5, the mass term and
the quartic coupling terms can both be added by divergence-free
perturbation series.

More directly, Monte Carlo studies may be made of the full nonlinear
lattice theory. This can be done with a variety of rescaled field
variables, but perhaps the most convenient is that of
Eq.~(\ref{h97}). In those field variables, the Euclidean  lattice
space probability distribution is given by \b &&C\s\exp\{ -\s\half\s
 {\t\sum_k} {\t\sum_{k^*}}\,(\phi_{k^*}-\phi_k)^2\s-\half [\s N'\s a\s q^{-1}\s]\s m^2\s
 {\t\sum_k}\s\p_k^2\no\\
&&\hskip1.27cm -[\s N'^{\s 3}\s a\s
q^{s-2}\s]\s\l\s{\t\sum_k}\s\p_k^4-\s\half \s
 {\t\sum_k}\EE_k(\p)\s\}\;, \e
 where $C$ denotes an overall normalization factor. Traditional
 Monte Carlo methods may be used with this weighting to approximately determine
 various correlation functions. The coefficients of the $m^2$ and
 $\l$ terms are direct transcriptions of those determined earlier in
 the present section. They have been estimated to be suitable for
 all values of $a$ and $L$ for which $L\s a<\infty$. Thus they
 should also hold in the continuum limit with finite spatial and spacetime volumes.
 Any divergences that arise in the infinite spatial and spacetime volume limit
 should only be those that typically arise for stationary questions,
 such
 as illustrated earlier with the one-dimensional Ornstein-Uhlenbeck
 process; such divergences are expected and do not require any special treatment.

\section{Additional Discussion and Conclusions}
In \cite{klau2} there is an extensive discussion of soluble, scalar
nonrenormalizable models that are idealized versions of the relativistic
model treated in the present paper. Such models differ from the
relativistic model in that in one case {\it all} spacetime
derivative terms are omitted from the classical action and in the
second case {\it all but one} of the derivative terms are dropped;
in this latter case, the remaining term is identified with the
eventual time direction in an analog of a Wick rotation. These
models have no physics and are only of academic interest.
Nevertheless, from a mathematical viewpoint both of these models
lead to Gaussian results if no further counter terms are introduced,
and if they are studied perturbatively, they are both
nonrenormalizable. Fortunately, both of these models have sufficient
symmetry so that they can be rigorously solved on the basis of
self-consistency without using any form of perturbation theory. One of the
results for both models is that as the coupling constant of the
nonlinear interaction term is reduced to zero, the theories do {\it
not} return to the appropriate free theory but instead they pass
continuously to an appropriate pseudofree theory. Moreover, both
theories exhibit meaningful, divergence-free perturbation theories
about the pseudofree theory, but definitely not about the customary
free theory. These models are explicitly worked out in Chaps.~9 and
10 of ref.~\cite{klau2}, but there is also a natural reason why
such results are plausible.

Nonrenormalizable quantum field theories have exceptionally strong
interaction terms. This statement can be quantified as follows:
Consider the $\varphi^p_n$ relativistic scalar theory which has a
free (Euclidean) action given by
  \b W =\half\tint[\s(\nabla \varphi)(x)^2+m^2\s\varphi(x)^2\s]\,d^n\!x\;, \e
where $x\in{\mathbb R}^n$. These theories have the nonlinear
interaction term
  \b V=\tint \varphi(x)^p\,d^n\!x\;, \e
  where we focus on cases where $p\in\{4,6,8,\ldots\}$. Such
  expressions appear in a formal functional integral such as
   \b S_\l(h)={\cal N}\int e^{\t \tint h\varphi\,d^nx-W-\l V}\,{\cal
   D}\varphi\;. \e
   If $\lim_{\l\ra0}\,S_\l(h)=S_0(h)$, then the interacting theory
   is continuously connected to the free theory; if, instead,
  $\lim_{\l\ra0}\,S_\l(h)={\tilde S}_0(h)\not= S_0(h)$, then the interacting
  theory is {\it not} continuously connected to the free theory, but
  rather it is continuously connected to a psuedofree theory. Under what conditions
  could this latter situation arise?

  Consider the classical Sobolev-type inequality (\cite{klau2}, Chap.~8)
  given (for $\varphi\not\equiv0$) by
    \b \{\tint \varphi(x)^p\,d^n\!x\}^{2/p}/\{\tint[\s(\nabla \varphi)(x)^2+
    m^2\s\varphi(x)^2\s]\,d^n\!x\}\le R \e
    as a function of the parameters $p$ and $n$. For $p\le 2n/(n-2)$, it follows that
    ${R}=4/3$; for $p> 2n/(n-2)$, ${R}=\infty$
    holds.
    This dichotomy is {\it exactly} that between
    perturbatively renormalizable and perturbatively
    nonrenormalizable models. But why should this inequality relate to
    renormalizability?

    It is the author's long-held belief that the explanation arises from a {\it hard-core behavior of
    nonrenormalizable interactions} \cite{kk6}. Simply stated, the interaction for
    such theories is so strong that a set of nonzero measure of the
    field histories allowed by the free theory alone is projected
    out when the interaction term is present. For example, ${R}=\infty$
    means that there are fields for which $V$ is not dominated by
    $W$ in the same way as when ${R}=4/3$. This fact suggests that
    for positive coupling constant values, some of the field histories are
    projected out never to return as the coupling constant passes to zero.
    This result is the effect of a hard core at work. For the
    first idealized model mentioned above, the analogous ratio is
     $\{\tint \varphi(x)^4\,d^n\!x\}^{1/2}/\{\tint m^2\s
     \varphi(x)^2\,d^n\!x\}$, which for any $n\ge1$ clearly has no finite upper
     bound, while for the second idealized model, the appropriate
     ratio reads $\{\tint \varphi(x)^4\,d^n\!x\}^{1/2}/\{\tint
     [{\dot\varphi}(x)^2+m^2\s\varphi(x)^2]\,d^n\!x\}$, which, in this case
     for any
      $n\ge2$, has no
     finite upper bound. These soluble models -- each more singular in principle
     than the relativistic models -- are examples of hard-core
     interactions that nevertheless have divergence-free
     perturbations about their own pseuofree model.
     This set of facts strongly suggests that
     relativistic scalar fields such as $\varphi^4_n$, $n\ge5$, as we
     have focused on, also have a corresponding hard-core behavior.

     The soluble, idealized models have formulations that involve
     inverse field powers; indeed, the second model, which lies
     closer to the relativistic models, has an inverse-square field
     power counter term in the lattice action itself. For both idealized models,
     the needed counter term was not assumed, it was {\it derived}, thanks to
     a large symmetry of the model. For the relativistic models, there is
     insufficient symmetry to derive the needed counter term, and thus the counter term
     must be postulated, i.e, guessed. There have been several past
     suggestions that have not lived up to expectations. The present
     paper offers one more proposal that seems to satisfy the
     expected requirements.

     Unfortunately, the present model is
     not (or at least seems not) analytically tractable, probably lacking
      a technical means to analytically perform perturbation calculations
     about the pseudofree theory.
     Nevertheless, it would seem
     possible that numerical Monte Carlo calculations should be
     feasible. The first such calculation that should be made is a
     test for non-triviality that is applied to such theories
     by testing whether or not the Gaussian property
     that
     \b \<\s(\Sigma_k h_k\p_k\,a^n)^4\s\>-3\s\<\s(\Sigma_k h_k\p_k\,a^n)^2\s\>^2=0 \label{k1}\e
     holds true for all choices of $\{h_k\}$ as one approaches the
     continuum limit; a single violation of this inequality would demonstrate that the
     continuum limit is not that of a free theory.
     In view of the connection of such full spacetime
     correlation functions to those on a single spatial surface, as
     shown in Sec.~4, it seems unlikely
     that (\ref{k1}) holds true thanks to the chosen form of the counter term;
     on the other hand, Sec.~4 ultimately
     involves inequalities which might allow (\ref{k1}) to sneak
     through.

     Should the above nontriviality test prove successful for a $\varphi^4_5$ relativistic
     model, for example,
     it would be worthwhile to study various three-dimensional models such as
     $\varphi^8_3$, $\varphi^{10}_3$, etc. {\bf [Remark:} It is noteworthy that if instead of
     the interaction $\l_0\, \p^4_n$ discussed in the present paper we had started
     with $g_0\, \p^{2r}_n$, where $r\in\{3,4,5,6,\dots\}$, then it follows that
     such an interaction would also possess a divergence-free perturbation
     expansion provided we choose
       \b g_0=N'^{(2r-1)}\,a^{(s-1)r-s}\,g\;, \e
       where $g$ is the physical coupling constant. Moreover, it follows
       that
          \b Z^r\,g_0=[N'\s (q\s a)^s]^{-1}\,g\;, \e
          for all values of $r$, just as was the case for the mass term in (\ref{x2})
          and for the quartic coupling in (\ref{x4}).{\bf ]} It would also
     be interesting to reconsider the $\varphi^4_4$ model, which,
     although perturbatively renormalizable, has the property of
     tending to a Gaussian theory in the continuum limit as shown by
     renormalization group studies as well as Monte Carlo studies.
     It is possible that a nonperturbative and
     nontrivial $\varphi^4_4$ model is still to be discovered.

     \subsubsection*{The counter term from a factor ordering ambiguity}
     It is not without interest that the chosen counter term can be
     viewed as arising from a factor ordering ambiguity. Let $\pi_k$
     denote the classical momentum conjugate to the field $\p_k$ for
     all $k$ in a spatial slice. The classical Hamiltonian reads
       \b H=\half{\t\sum}'_k\s \pi^2_k\, a^s+{V}_0(\p)=\half{\t\sum}'_k D\s\pi_k\s
       D^{-2}\s\pi_k\s D\, a^s+{V}_0(\p)\;, \e
       where $D=D(\p)\equiv\Pi'_k[\s\S_l
       J_{k,l}\s\p_l^2\s]^{(N'-1)/4N'}$.
       Passing to the quantum theory leads to the Hamiltonian
       operator [c.f., (\ref{j14})]
         \b &&\H=-\frac{\hbar^2}{2}\s a^{-s}{\t\sum}'_k D\frac{\d}{\d
       \p_k}\s D^{-2}\s\frac{\d}{\d\p_k}\s D+{\cal V}_0(\p)
       \no\\ &&\hskip.48cm
       =-\frac{\hbar^2}{2}\s a^{-s}{\t\sum}'_k\frac{\d^2}{\d\p^2_k}+{\cal
       V}_0(\p)+\frac{\hbar^2}{2}{\t\sum}'_k\F_k(\p)\,a^s\;.\e

     \subsubsection*{Classical limit}
     When dealing with a nonrenormalizable $\varphi^4_n$ theory, we argued
     in  Sec.~1 against choosing either no counter terms or those counter
     terms
     suggested by a regularized perturbation analysis about the free
     theory. This was due, in part, to the fact that such theories tend not to have the
     correct classical limit, namely, the original nonlinear
     classical theory one started with. That property is clear in
     the case of no counter terms, which leads to a free quantum theory,
      and it is effective as well when perturbative
     counter terms are considered because in the latter case there
     is no complete and well-defined quantum theory for which the classical limit can be studied.

     One strong test of whether or not the ideas in this paper have
     some validity would be to try to take the classical limit and
     confirm that the expected nonlinear relativistic theory
     emerges. The study of this question first requires having some
     control on the continuum limit, but  in support of its possible realization we note that the
     second idealized model treated in Chap.~10 of \cite{klau2},
     namely, the model including the time derivative of the field,
     has been shown to indeed have the correct classical limit for
     the idealized but nonlinearly interacting model in question.
     It is possible that similar techniques may be used to show that
     the presently proposed quantization scheme for
     nonrenormalizable $\varphi^4_n$ models has a classical limit
     that agrees with the original nonlinear classical theory.

     Even if our proposal leads to a nontrivial quantum theory, and
     even if that quantum theory exhibits a correct classical limit,
     the question may arise whether this proposal for quantization
     is the ``correct'' quantization procedure. As in any
     quantization procedure, where one starts from a theory with
     $\hbar=0$ and constructs a theory with $\hbar>0$, there is a
     great deal of latitude in the result. Nevertheless, in the
     absence of any other satisfactory proposal to deal with
     nonrenormalizable theories, one might look favorably on a model
     that offers more than was previously available.

\end{document}